\documentclass[conference]{IEEEtran}
\IEEEoverridecommandlockouts
\usepackage{cite}
\usepackage{amsmath,amssymb,amsfonts}
\usepackage{algorithmic}
\usepackage{graphicx}
\usepackage{textcomp}
\usepackage{xcolor}

\def\BibTeX{{\rm B\kern-.05em{\sc i\kern-.025em b}\kern-.08em
    T\kern-.1667em\lower.7ex\hbox{E}\kern-.125emX}}
\begin{document}

\title{Pitako - Recommending Game Design Elements in Cicero\\
}

\author{\IEEEauthorblockN{Tiago Machado}
\IEEEauthorblockA{\textit{Game Innovation Lab} \\
\textit{New York University}\\
New York, USA \\
tiago.machado@nyu.edu}
\and
\IEEEauthorblockN{Dan Gopstein}
\IEEEauthorblockA{\textit{Game Innovation Lab} \\
\textit{New York University}\\
New York, USA \\
dgopstein@nyu.edu}
\and
\IEEEauthorblockN{Andy Nealen}
\IEEEauthorblockA{\textit{School of Cinematic Arts} \\
\textit{University of Southern California}\\
Los Angeles, USA \\
andy@nealen.net}
\and
\IEEEauthorblockN{Julian Togelius}
\IEEEauthorblockA{\textit{Game Innovation Lab} \\
\textit{New York University}\\
New York, USA \\
julian@togelius.com}
}

\IEEEpubid{\begin{minipage}{\textwidth}\ \\[12pt]
978-1-7281-1884-0/19/\$31.00 \copyright 2019 IEEE
\end{minipage}}

\maketitle

\begin{abstract}
Recommender Systems are widely and successfully applied in e-commerce. Could they be used for design? In this paper, we introduce Pitako\footnote{\textit{The name Pitako comes from the Portuguese dialect word - pitaco - in use around the Pernambuco state in Brazil to designate hints coming from people who supposedly know what they're talking about.}}, a tool that applies the Recommender System concept to assist humans in creative tasks. More specifically, Pitako provides suggestions by taking games designed by humans as inputs, and recommends mechanics and dynamics as outputs. Pitako is implemented as a new system within the mixed-initiative AI-based Game Design Assistant, Cicero. This paper discusses the motivation behind the implementation of Pitako as well as its technical details and presents usage examples. We believe that Pitako can influence the use of recommender systems to help humans in their daily tasks.
\end{abstract}


\section{Introduction}
Recommender Systems are most well-known for their usage in e-commerce. The number of algorithms, applications, and studies is so large that some scientists affirm we are living in the age of recommender systems \cite{anderson2004long}. Their techniques are nowadays used in many domains, from movies \cite{Said:2011:PTF:2039320.2039328} and books \cite{Pera:2014:WHO:2682648.2682787}, to scientific papers \cite{SesagiriRaamkumar:2018:MES:3213586.3226215}, fitness training \cite{TurmoVidal:2018:MCI:3196709.3196789} and even friends \cite{Liu:2014:RFL:2640087.2644197}, it seems everything can be suggested by these tools. 
However, we have not seen much use of recommender systems to assist humans in design tasks. Specifically, if we consider the field of AI Game Design Assistants, despite the increasing number of level generators and telemetry tools available, it is not easy to find a human-in-the-loop process in which a machine is constantly evaluating and providing suggestions about what to do next in the game design process. 
One example of a recommender system being used for game design is Sentient Sketchbook \cite{liapis2013sentient}. Here, evolutionary algorithms are used to create level suggestions based on an initial user level given as input. 
Still, it is limited to the level design of a predefined game genre and the generated suggestions are not based on any existing content designed by other humans, so that system fails to build on existing designs.
In this paper, we present Pitako, a recommender system to assist game designers in the design of the games themselves. Pitako suggests mechanics and dynamics to add to games based on association rules and frequent item sets found in other games across similar and different genres.
The motivation to design such a system comes from the observation that many games are built upon features borrowed from their predecessors. Jesper Juul illustrated this phenomenon as a network of common mechanics available in matching tile puzzle games, and a family tree outlining how these mechanics have spread throughout games \cite{juul2007swap}. Similarly, the work of Summerville et. al. \cite{Summerville:2017:MAR:3102071.3102104} analyzes the jump mechanic available in 2D platform games developed for the Nintendo Entertainment System (NES). The work presents commonalities and trends in jumping across games, developers, and game franchise.
This motivated us to investigate a system that could automatically do the same sort of taxonomy research introduced in Juul's paper and the formal categorization presented in the Summerville's one to describe mechanics in games similar to the ones being developed.
The idea of doing exploratory design in an automatic way also brings the possibility of exploring such a tool as an educational environment. 
Our proposal relies on the fact that Pitako performs an automatic exploration of potential design elements and provides them to the user, ready to be incorporated into the game they are designing. Such flexibility allow the designer to inspect, play, analyze, and learn alternatives to their games in a way that couldn't be possible without the use of an AI-assistant.
Therefore our contribution is providing creative stimulus since the user can take the recommendations as an inspirational tool for generating new ideas, as mentioned by Shneiderman\cite{shneiderman2000creating}. Plus, we advocate the use of recommender systems for assisting human tasks, and we hope this work can inspire and push other researchers to go beyond of the e-commerce-based paradigm.

\section{Background}

\subsection{The atoms of a game}
Digital games as we know them today are a creation of the twentieth century. As every invention of the 1900s the analysis (and, in some cases, the design) of the digital games was influenced by the three main scientific discoveries that started to change the world more than a hundred years ago: the atom, the gene, and the byte (or bit) \cite{siddharta2016gene}. Conceptually speaking, these three discoveries brought the evidence of the unit, to the indivisible and smallest part of a system as the main component to understand the whole.
From the game literature, we can see that this idea is widely shared. For example, the idea of the game as a sum of its units is presented by \cite{sicart2008defining}, and minimalist game design \cite{nealen2011towards} advocates a development practice in which
you start your new game project from its core game mechanics. Everything else
like graphics, sounds and even new mechanics (derived from the basics 
one or not) comes after. In the book written by Brenda Brathwaite and Ian Schreiber \cite{brathwaite2008challenges}, a best seller in the game design literature, 
you can find a whole chapter entitled "game design atoms". The authors claim that ``by looking at a game as a collection of atoms (mechanics, dynamics, goals, game state, etc.), the process of design itself becomes clearer". A similar approach is put forth in another best seller, written by Paul Schuytema \cite{schuytema2007game}. The MDA - Mechanics, Dynamics, and Aesthetics \cite{hunicke2004mda} also brings the same concept of building games starting by its basic components, and evolve them until the level of creating the art content(aesthetics). 

\subsection{The catalog}
To facilitate the process of knowing, recommending and combine features from different games to design new ones, we took inspiration from chemical and biological methods. For example, the geneticist Thomas Morgan designed a catalog of fruit flies in which he described 
their minimal features and tracked every mutation to facilitate how to combine them and come up with modified species \cite{morgan1925genetics}. Also, 
the chemical periodic table of elements was designed to make easier the process of synthesizing chemical structures.  Therefore we build a catalog of games, also based on their minimal components with the intention of inspect their individualities and mix them to generate new games.

\subsection{VGDL \& GVGAI}

\begin{figure}[!t]
\centering
    \includegraphics[width=1.0 \columnwidth]{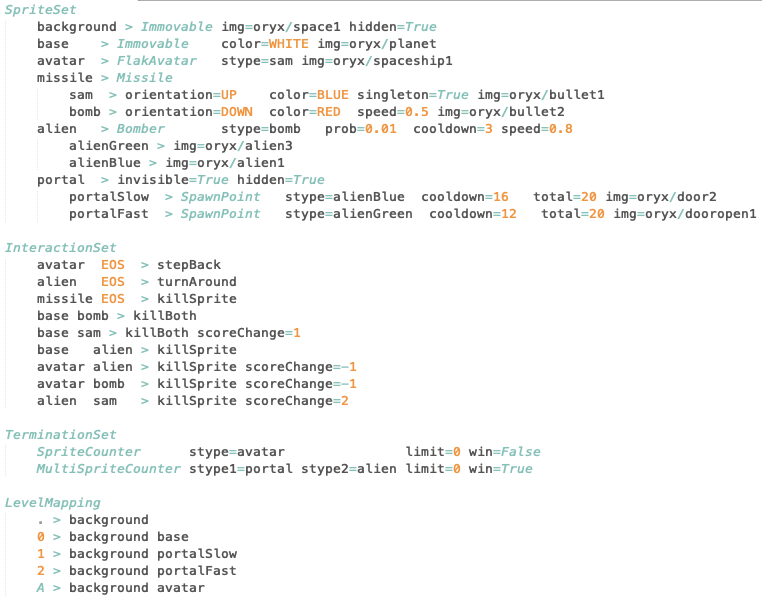}
    \caption{Example of a \textit{Space Invaders (Taito, 1978)} version written in VGDL.}
\label{vgdl}
\end{figure}

To implement a design breakdown process and build a catalog of games and their elements, we used a Game Description Language (GDL) that could be as atomic as possible and at the same time as universal as possible.
The Video Game Description Language (VGDL)\cite{ebner2013towards, schaul2013video} is a domain-specific language for general AI
game design. It allows the design of many game genres such as puzzles, action, shooters, and so on. Graphically the games are similar to the style of Atari 2600 or Nintendo Entertainment System (NES) games. They are all two dimensional and constrained to grid-based level design.
VGDL was created primarily for research in general video game playing, and the primary design consideration was the suitability for testing AI agents.
Therefore, you can with the same description language have a puzzle labyrinth like \textit{Pacman (Namco \& Atari, 1980)} and a shooter like \textit{Space Invaders (Taito, 1978)}, both of which can be easily played by autonomous systems.
The VGDL description of a game is short and human readable (See Figure \ref{vgdl}). It is defined by four sets. However we will just explain with more details the ones most important for this study.

\begin{itemize}
    \item \textbf{The Sprite Set} - A sprite is any object in the game, including its graphical representation and behavior. In this set the sprites are defined. It is the place to specify if your avatar can shoot and if a non player character (NPC) will move randomly around the level or chase another game element. We stress here that in VGDL a sprite has both a graphical interpretation and a behavior associated with it. This is essential for this system since it is one of the main components of a mechanic rule system in VGDL. Therefore, every time the word sprite appears in the text, it is related to the concept of a sprite in VGDL.
    
    \item \textbf{The Interaction Set} - This set defines what happens when two sprites collide. It is the second part of the mechanic rule set in VGDL. Here, for example, you define when an element should be removed from the game after another one hits it. For example, when a missile collides with an alien airship and both sprites disappear.
    
    
\end{itemize}

The VGDL description still contains a \textbf{LevelMapping set} that maps sprites into symbols to represent them in the game level matrix, a \textbf{Termination Set} that defines game over conditions for winning and losing the game, and finally the \textbf{levels}, these are designed in a 2D matrix of symbols representing the places the sprites will occupy in the level when the game starts.
The General Video Game AI (GVGAI) is a framework for general AI video game playing \cite{perez2016general, perez20162014}. It has an associated competition, which runs annually. Participants submit their agents which are judged by their performance playing unseen games. All the games available in the framework are coded in VGDL, some of them are famous versions of classics like \textit{Sokoban (Thinking Rabbit, 1982)} or \textit{Zelda (Nintendo)} cave levels. 

\subsection{Game design breakdown}

Before recommending fundamental components for the design of a new game, we must first develop a comprehensive understanding of similar games, from which to draw patterns to be recommended. We call the process of understanding the fundamental components of similar games "game design breakdown". It consists of getting a formal game
description and rewriting it in a simplified way to have access to its minimal (atomic)
elements. The process' name comes from a similar process in the movie 
industry, the script breakdown \cite{singleton1991film}, in which the crew reads the movie 
screenplay page by page and define all the minimal technical aspects necessary to shoot the movie. To keep using our atomic metaphor, we took the 
individual sprite and sprite-interaction descriptions as our atomic elements. Everything they bring with them 
(their parameters) are the subatomic elements. It allow us to build a data set where we can map games to the elements they contain, and sub-elements to their parent elements. 
\subsection{Cicero}
Cicero is an AI game design tool that offers several kinds of assistance. It has a replay analysis tool\cite{Machado:2017:SRA:3102071.3102090}, a query engine\cite{machado2019kwiri}, a level visualization interface, and a game-debugger assistant\cite{Machado2018AIAssistedGD}. It is built on top of the GVGAI framework and runs VGDL games, which allows Cicero to automate gameplay by interacting with games using one of the many agents available. There are agents based on various techniques, including Monte Carlo Tree Search, evolutionary planning and reinforcement learning. In essence, any kind of algorithm that one can design or adapt for playing games is supported by GVGAI, and consequently by Cicero. This feature was used to design an AI debugging experiment, which showed that users improve their accuracy during game debugging tasks when augmented with AI-assistance.
Cicero also has a pre-existing recommender system for choosing sprites based on analysis of the game similarity through Euler vector distances. The sprite attributes are written as vectors and compared in different games. A score ranks the most similar ones and suggests them\cite{machado2016shopping}.
Unfortunately, this approach does not reach good results.
Attributes are not well differentiated from one sprite to another. Recommendations are often repeated, and there is no information provided about why an element is being recommended.


\subsection{Recommender systems and the Apriori algorithm}
\begin{figure}[t!]
\centering
    \includegraphics[width=0.7 \columnwidth]{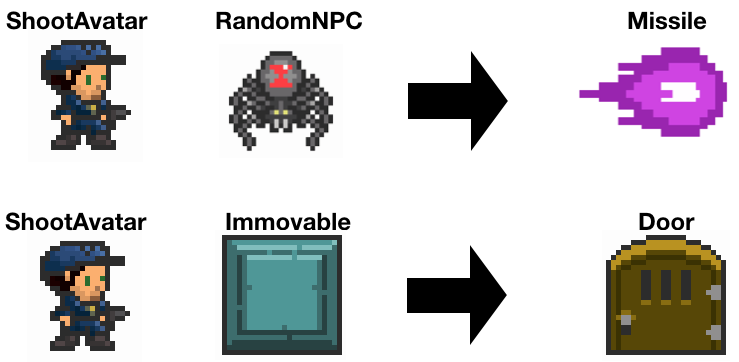}
    \caption{In the association rule on the top, the presence of a \textbf{ShootAvatar} and a \textbf{RandomNPC}, in a sprite set, implies in the presence of the \textbf{Missile} sprite. In the association at the bottom, the presence of a \textbf{ShootAvatar} and an \textbf{Immovable} implies in the presence of the \textbf{Door} element.}
\label{fig:assoc}
\end{figure}

A popular definition of `recommender system' is ``any system that produces individualized recommendations
as output or has the effect of guiding the user in a personalized way to interesting
or useful objects in a large space of possible options. Such systems have an obvious
appeal in an environment where the amount of on-line information vastly outstrips
any individual’s capability to survey it."  \cite{burke2002hybrid}.
In order to attain the goal of providing accurate recommendations, a myriad of techniques are available\cite{Adomavicius:2005:TNG:1070611.1070751},  from `authoritativeness" criteria \cite{brin1998anatomy} to collaborative filtering, what is probably the most popular and mature recommendation technique in use now \cite{billsus2000user, schwab2001learning}, and largely used by companies such as Amazon \cite{linden2003amazon} and Netflix \cite{gomez2016netflix}. These techniques present strengths and weakness and its common to hybridize them to overcome the issues.
In our case, we decided to use association rule mining for recommender systems \cite{lin2002efficient}, more specifically the apriori algorithm \cite{agarwal1994fast} because the way it creates rules based on a set of transactions suits the data we can extract from VGDL game descriptions. The apriori algorithm efficiently analyzes a database of transactions to find association rules between repeated items in different transactions. An association rule says that the presence of an item \textbf{X} in a transaction implies in the presence of \textbf{Y} in the same transaction with some probability (see Figure \ref{fig:assoc}). A common example of this technique is the supermarket basket case. The article published by Agrawal et. al. \cite{agrawal1993mining} exemplifies by giving the statement that 90\% of transactions that purchase bread and butter also purchase milk. The presence of bread and butter in a transaction implies the presence of milk in the same one. 90\% is the confidence that such a rule will hold in a future set of new transactions. Finding all such rules is valuable for cross-marketing and attached mailing applications. Other applications include catalog design, add-on sales, store layout, and customer segmentation based on buying
patterns \cite{agarwal1994fast}.
We believe that games share behaviors in the same way that shoppers share items in their shopping carts. For example, \textit{Mega Man X (Capcom, 1993)} and \textit{Super Metroid (Nintendo, 1994)} have the same mechanic of shooting. On top of that, they even share the ``hold the shoot button'' rule to fire a more powerful shot. The idea here is that if your game has a shooting mechanic you may want to use a powerful shot rule like the cited games. The presence of the shooting mechanic implies (with a certain probability) in the presence of the powerful shot rule. Given it is already available in previously designed games, it can be directly imported rather than re-implemented.
Our system also displays confidence levels along side each of its suggestions to help the designer understand the strength of each recommendation. Commonly used mechanics like the `shooting example' above will appear with a high confidence level. Less common mechanics will have a low confidence level. This way, the designers can decide in which direction they will push their new games. Do they want to make a clone of an existing popular title (or maybe learn how it was done) ? Then, they can get the high confidence suggestions. Do they want to come up with something new and explore different possibilities? If so, they can go for the low confidence recommendations. Of course, they can use a mix of high and low confidence game elements, and mix them with the ones they are creating by themselves.

\subsection{Suggestion engines for design assistance}
While recommender systems are predominantly used in the e-commerce sector there are examples in the literature of other sorts of suggestion engines. They do not necessarily rely entirely on recommender systems techniques, but are an inspiration to our work because they share the idea of searching for patterns and suggesting them to assist design tasks. 
In the field of sketch based interfaces, Igarashi et. al. \cite{igarashi2001suggestive} introduced a system that assists humans in 3D modeling works. When users sketch an object, the 3D structure is compared against a database of previously sketched structures. When there is a match between the structures, the parts yet to be modeled are suggested by the system and the user may choose to utilize one.
A paper published by O'Donovan et. al. \cite{O'Donovan:2015:DDI:2702123.2702149} describes the design of a virtual illustrator assistant. 
In this work, the users can design layouts for business cards or birthday invitations. At the same time, the system is generating altered copies of the users' original work. These copies change spacing, text font, add and remove figures, backgrounds, resolution and so on. This way, the authors claim that the users benefit from an automatically exploratory design task, which they promote as a vital part of the design process.
In the work by Nguyen et. al. \cite{Nguyen:2015:KSS:2702123.2702411} the authors bring a recommender system for topic conversation suggestions. The idea is to provide a way of two strangers to engage in a good conversation easily. The system takes in consideration a list of subjects that the users have in common and provide topics within them.  The  results contributed to the understanding of how communication interventions influence people's experience and behaviors, and enhance interpersonal interactions.
The (already cited) Sentient Sketchbook brings a mixed-initiative method based on evolutionary techniques for assisting the level design of real-time strategy games by providing suggestions based on the level being designed by the user. Finally, Evolutionary Dungeon Designer is also based on evolutionary methods. It also applies heuristics based on game design patterns to assist users in the creation of dungeon maps for adventure games. The tool follows an interaction design approach with every new verision based on previous users' feedback\cite{Baldwin:2017:TPM:3102071.3110572, Alvarez:2018:FCM:3235765.3235815}.
All of theses examples show that it is possible to use recommender system techniques to assist human tasks. 
However, it seems that its usage in the game domain just started to scratch the surface. 

\section{Recommending game design elements}
In this section we show how three types of game elements are recommended: sprites, sprite placements, and interaction rules. It is worth to note that the confidence level of the suggestions in this particular case of a recommender system has a different interpretation. A low confidence level or a high one does not mean a bad or good item (sprite, interaction or position) to be picked. It works here as a guide to the designers. If they want to let their games as close as possible to the original source they will choose the suggestions with high confidence levels. If they want to differentiate their games, they will choose the recommendations with low levels of confidence. 

\subsection{Apriori algorithm applied to sprite sets}
\begin{figure}[h]
\centering
    \includegraphics[width=0.7 \columnwidth]{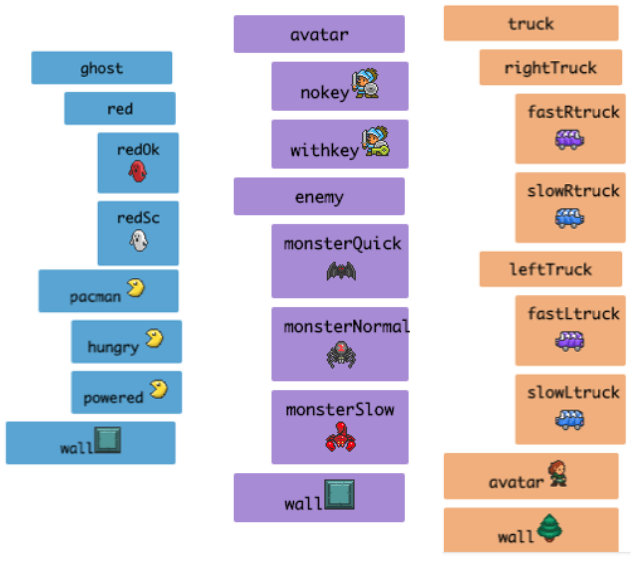}
    \caption{From left to right, a catalog of game sprite sets with \textit{Pacman (Namco \& Atari, 1980)}, \textit{Zelda (Nintendo)} and \textit{Frogger (Konami, 1981)}. The games are arranged as a list of "baskets" (in Apriori terminology) where each game is a basket and the elements (sprites) are items. This approach makes easier the process of searching an element and combine it with one (s) from other games.}
\label{fig:catalog}
\end{figure}

As stated before, a sprite in VGDL is not only an image. It is the description of a behavior onto which you can attach an image. If you are implementing the hero of a shooter game like \textit{Mega Man X (Capcom, 1991)} or \textit{Contra 3 (Konami, 1992)}, you would specify the sprite behavior as a ShootAvatar, and then attach the image of your hero to the defined sprite. Therefore, the set of sprites is the first source of rules in a VGDL game, and a natural candidate to be automatically recommended during AI-assisted game design.
To use the apriori algorithm to provide sprite suggestions the first step is collecting every sprite in every known game in our catalog (Figure \ref{fig:catalog}). In the end, we have a table where each game represents a transaction set, with the sprite behaviors as the items of the transactions.
The final table is therefore the algorithm input. After running apriori we get a table with the item sets whose frequency of occurrences reach the limit of the minimum support provided. Therefore, the table shows us how frequent sprite behaviors are associated with each other in different games.

\subsubsection{Sprite Recommendation protocol}
\begin{figure}[h]
\centering
    \includegraphics[width=1.0 \columnwidth]{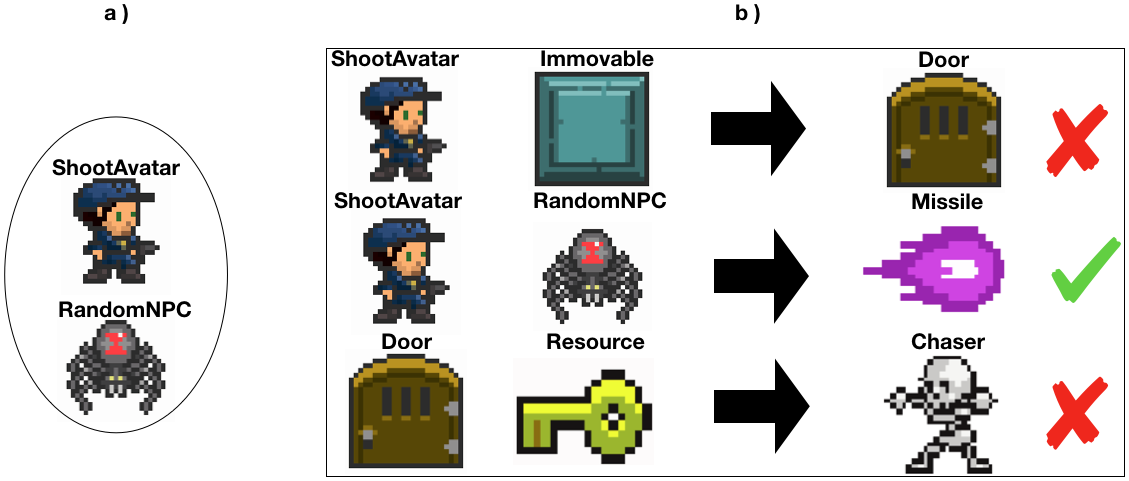}
    \caption{In this figure (a) is the user's game's sprite set. In (b) we look at the association set, to find an association rule related to (a). The selected association rule, will give us the candidate sprite to be suggested. We then, go to the games in the catalog to find one with the given sprite. Finally we copy and recommend it to user. }
\label{fig:proto}
\end{figure}

With knowledge of the association between different sprites in different games, we can then provide suggestions to the designer. The input for our recommendation protocol is the sprite set of the game in development. This set is compared against the set of association rules. All the associations which have the sprite set as a subset are stored in a list. At this point, the association rules give us the sprites candidates to be recommended ( Figure \ref{fig:proto}). In this approach sprites from the same type can be part of the recommendation list. The user can use filtering commands to avoid ``repetitions" when consulting the list. With the sprite candidates pointed by the association rules, we go to the catalog and find which games have the same sprites. We then do copies of these sprites and suggest them to be used in the user's game. Note that a sprite cannot be suggested direct from an association, because it is necessary to adapt contextual information from the game which it is recommended, and the one which is getting the sprite. We talk more about this process in the following section. 
\subsubsection{Game Blending Process}

Once the user has selected which sprite to add to the game being developed, we start the procedure to merge the features and behaviors of the sprite into the game; we call this by the blending process, and we are inspired by the logical blend methods present in the work by Eppe et. al. \cite{eppe2018computational}. We are extracting information from one game and inserting it into another one. As sprites can have other sprites attached to it, we need to make sure that it will be imported not individually, but with all the other sprites it refers.
We use this approach to reduce user workload. For example, if a designer selects a shooter character, the recommender will import the character and the object it uses to shoot. Otherwise, the designer would have the shooter character but would have the extra work to design an object to be shot.
In order to create the sprite package to be imported, we use a depth-first search tree navigation to navigate in the game catalog entry recursively, so we can get the sprite and any other it refers to. It captures the context of a game and let it logically prepared to be inserted into another one. 


\subsection{Apriori algorithm applied to the sprite placement recommendations}
Once we have the user decision about the imported sprite, we can suggest positions in the game level map to put the recommended sprite. This process uses a combination of heuristics and the apriori algorithm.

The recommended sprite comes from a single previously designed game. In the VGDL game repository, each game has five levels. The input table for the apriori algorithm is generated by picking the recommended sprite type. The Sprite positions are then stored and indexed by level. So each level of the game is an association set of the positions of a specific sprite. The positions are the items of the association. 
Then, after we run the apriori algorithm, the output is a set of the most frequent positions for a specific sprite.\\

\subsubsection{Heuristics for placing sprites in a level}
After getting the output of the apriori algorithm with the positions to suggest, we apply filters to provide better recommendations to the user. The use of game level design heuristics is useful for this task.

\begin{itemize}
    
\item \textbf{Items should be at a K distance of the avatar}	\\We identified that the avatar should begin each level somewhat isolated from other sprites. For example, it is often undesirable to place a sprite adjacent to the avatar as the two sprites may collide on the first timestep, potentially harming or destroying the avatar as soon as the game starts. If the recommended sprite is a collectible item or the final goal of the level, placing it too close to the avatar will afford no challenge in retrieving it. By contrary, sprites which are harmful to the avatar are suggested to be placed close to resource items and exit level doors.
The interaction set provides us the context to understand which sprites can be harmful to the avatar player.
\end{itemize}
\begin{itemize}

\item \textbf{Ensure that the items will be placed inside the level boundaries} 
\\As the level being developed can have different dimensions from the level of other games in the catalog, positions of a specific sprite type that cannot fit in the user's level boundaries are not suggested.
\end{itemize}

\begin{figure}[h]
\centering
    \includegraphics[width=1.0 \columnwidth]{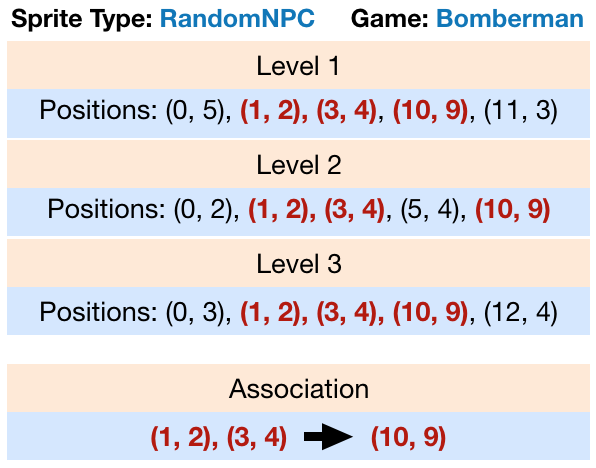}
    \caption{For the sprite type \textbf{RandomNPC} in the game \textbf{Bomberman}, its positions in the three first levels are shown in this figure. The association rule is filtering according the heuristics and then suggested to the user.}
\label{fig:rec_pos}
\end{figure}

After applying the heuristics and filtering the output of the apriori algorithm output (Figure \ref{fig:rec_pos}), the placement suggestions can then be recommended.

\subsection{Recommending interaction rules}
In VGDL, an interaction is composed by two sprites and the interaction they activate when colliding.
Therefore to recommend interactions, first it is necessary to do a map of combinations between all the sprite types (the key) and all the kind of interactions (the value(s)) they can fire when a collision event happens. Secondly, combinations among the elements of the sprite set in development are generated. Finally, a search identifies which combinations in
the user's game has an interaction that can be used as suggestion.\\  

\subsubsection{Sprite combination map}
To generate all combinations of sprites, it is necessary to navigate through the interaction set of all the games in the catalog, build a pair of sprite types and the interaction that is activated when they overlap each other. It's important to note that pairs of sprites with the same type are valid since often two sprites of the same type can interact with each other. The end of this process is a map whose key is the pair of sprites and the value is a list of the interactions activated by them.\\

\subsubsection{Sprite pairs comparison}
We can only recommend interactions for pairs of sprites available in the sprite set of the game in development. Therefore, after combining these sprites we compare each one of them against the map keys. When we found an entry whose pair (key) is equal to the pair in the user's game, we return the key value, a list of interactions. 
These are the ones that will be suggested to the user. In opposite to the sprite recommendations, interaction ones are straightforward. Therefore, there is no necessity of performing extra navigation in the game catalog entry, once the user decides for picking one of the recommendations, it is automatically inserted in the user's game interaction set (and removed from the recommendation set).

\section{Recommender System UI}
For the user interface design of our system, we divided it in three different Graphical User Interfaces (GUIs), one for the sprite set, another for the interaction set, and finally the level layout design to show the recommended positions of where to place the sprites.

\begin{figure}[!t]
    \begin{center}$
    \begin{array}{cc}
    \includegraphics[width=34.5mm]{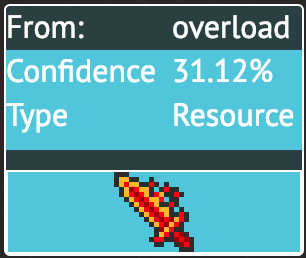}&
    \includegraphics[width=30.3mm]{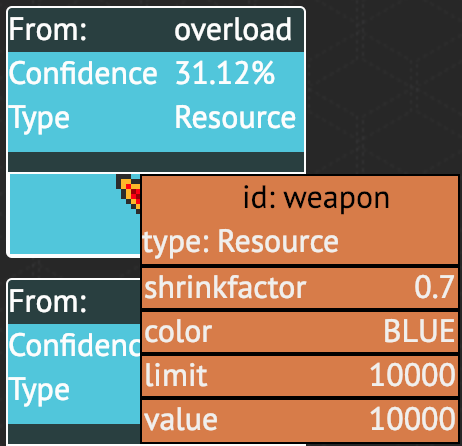}
    \end{array}$
    \end{center}
    \caption{The sprite recommendation component displays which game the sprite is imported from, the confidence of the recommendation, and its type (left figure). By hovering over it, the user can see other attributes and values of the sprite (right figure).}
    \label{sprite_recommender}
    \end{figure}
    
\subsection{Sprite Recommender UI}
The sprite recommendations appear on the screen as a list sorted in decreasing order by the confidence of the recommendation. Each list object has information about the game from where the recommendation is coming, the confidence value of the recommendation, the type of the sprite being recommended, and its image (Figure \ref{sprite_recommender}). By hovering over the image, a pop up shows a list of the sprite's attributes (and their values). We added this extra information to help the user to decide what to pick when having two or more sprites from the same type being recommended. When the user selects a recommended sprite, it is imported to the game sprite set. In the background, the system performs a search in the catalog, when finding the game whose specific sprite is recommended, it extracts all the information from the sprite and others it may carries. For example, if a user is picking a sprite which can shoot, the sprite that represents the projectile behavior is imported as well. 

\subsection{Interaction Recommendation UI}
\begin{figure}[h]
\centering
    \includegraphics[width=1.0 \columnwidth]{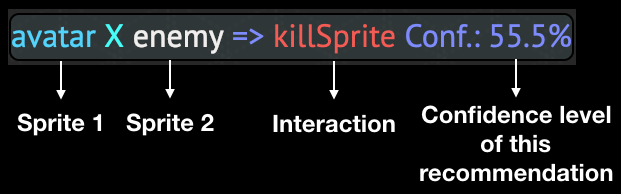}
    \caption{An example of a interaction recommendation component. The avatar sprite will be killed (killSprite interaction event) when colliding with a sprite enemy.}
\label{fig:interaction_ui}
\end{figure}

Similar to the sprite recommendations, the interaction recommendations (Figure \ref{fig:interaction_ui}) shows up as a list sorted by confidence in decreasing order. As the number of interaction suggestions tends to be high, the users can sort the list based on the elements they want to put to activate an specific interaction. When the user selects one the recommendations it is added to the interaction set. As mentioned before, interactions are plain descriptions of what happens in the game when a sprite A overlaps a sprite B. Therefore, this is, in general, a straightforward import process and extra search steps usually are not necessary because interactions do not have attributes which are other interactions (like a sprite having another sprite on its attribute list). It may have references for another sprite in particular cases. For example, when a sprite \textbf{X} transforms into a sprite \textbf{Z} after colliding with a sprite \textbf{Y}. In cases like this the system will import \textbf{Z} and all the other sprites it refers to the game. 

\subsection{Sprite Placement UI}
\begin{figure}[h]
\centering
    \includegraphics[width=.9 \columnwidth]{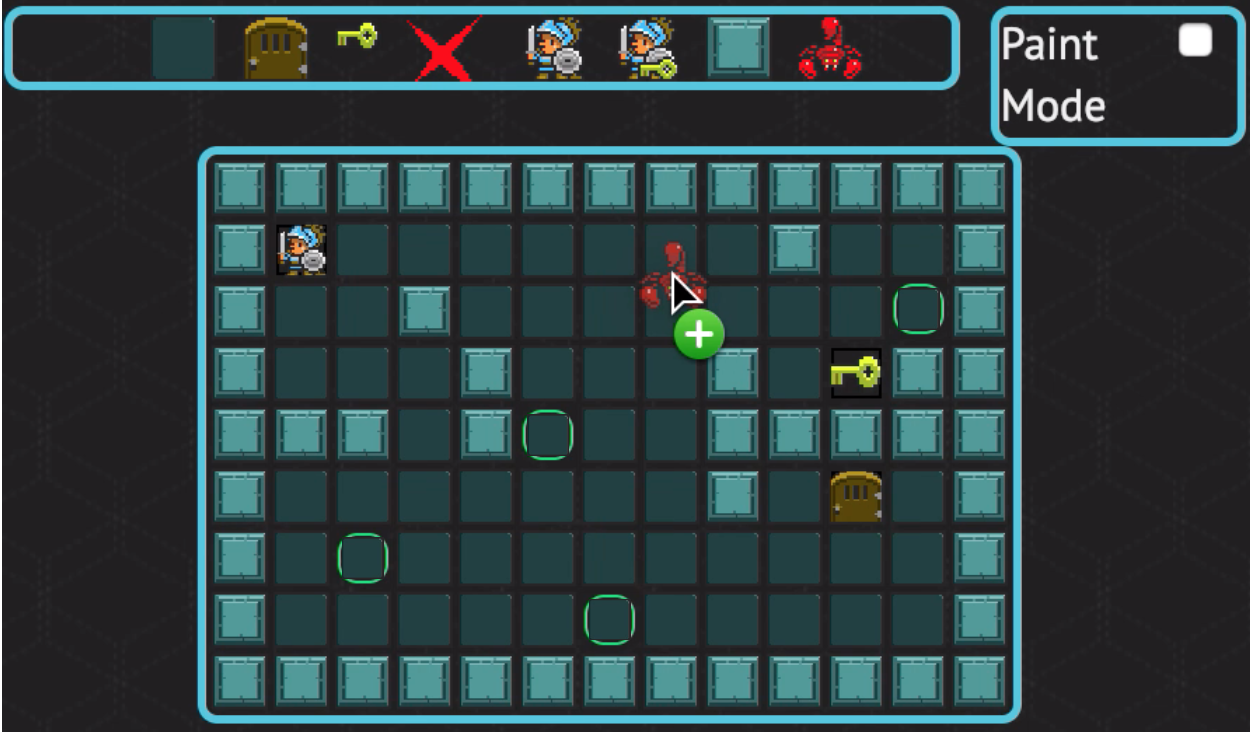}
    \caption{An example of recommending placement positions to a sprite. It is suggested, in this case, to place the enemy sprite far from the avatar player (top left corner) to avoid an instant kill when the game starts.}
\label{fig:drag_ui}
\end{figure}

The sprite placement suggestion happens whenever a user is adding a sprite in the level grid layout (Figure \ref{fig:drag_ui}). The process of placing a sprite in the grid is a simple drag-and-drop interaction. Therefore, as soon as the user starts to drag the sprite, green circles highlights the grid square positions recommended to receive the sprite.

\section{A sample session}
\begin{figure}[h]
\centering
    \includegraphics[width=1.0 \columnwidth]{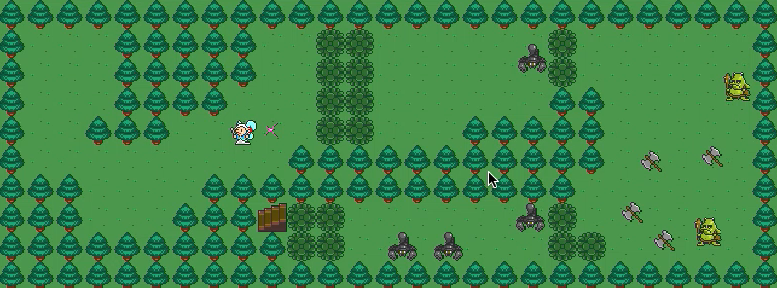}
    \caption{An example of a game designed by a user assisted by our recommender system. This game combines elements from \textit{Plants vs. Zombies (PopCap Games)} and \textit{R-Type (Irem, 1991)}}
\label{fig:game}
\end{figure}

The version of the system introduced in this paper was used for an informal test session. A computer science graduate student with experience
in game development, and a digital media professional, also with experience in game design tested the tool.   
We started their sessions by introducing the system and teaching them how to add sprites and interactions without recommender assistance. Then, we teach them how to use the recommendations. After the tutorial was finished, we asked them to use the tool freely and design a prototype of a hack-and-slash game in the style of the ones available for the Super Nintendo Entertainment System, like \textit{Knights of The Round (Capcom, 1991)}. 
In both of the experiences the recommender system was  used as expected.
Half or more of the sprite set was composed by recommended sprites. Users especially benefited from recommendations of sprites able to cast or spawn objects. In most of the cases they accepted the sprites (and the ones attached to them) exactly as they were suggested, in other cases they benefited from the customizing available and made some adjustments to fit the sprites behavior to what they had in mind for the game. In both situations the users interactions were faster than having to design the equivalent functionality from scratch. This is evidence that we can prove with Hierarchical Task Analysis (HTA). For having a \textit{FlakAvatar} in the game, by using Pitako the user will save, at least, three UI interaction steps. More than that, they indicated the suggestions gave them ideas about design elements they were not thinking before. For the interactions, almost half of the recommendations were accepted. 
 The number of recommendations accepted were far higher than what we expected. We can say that half or more of the game design happened in a mixed initiative way, with a human accepting the suggestions of an algorithm, and in some cases, making modifications to fit their needs. More than that, the users really put in effort and didn't only choose the first recommendations they were given. They really took their time to understand the recommendations, to inspect and test them, to change what they wanted to change, all based on a desire to make the best possible game in tandem with the AI-assistant (Figure \ref{fig:game}).

\section{Conclusions and future work}
In this paper, we discussed the design and implementation
of Pitako, a recommender system for game mechanics. We
took inspiration from previous works which found similarities
in games. These similarities were sometimes related to a
whole game genre, like puzzles that have the same mechanics
as Tetris [8] or a single mechanic feature like jumps in
2D platform games[9]. We developed our  system on top
of the VGDL and GVGAI framework to be  general across
game genres and mechanics. First, we reduced all the game
descriptions available to a description in which we could have
easy access to all the game components,  like a catalog of
chemical elements or biological species. The recommender
system itself is based on association rules and frequent item
sets. By applying the apriori algorithm,  our search process
navigates into our catalog of game elements and finds associ-
ations between the sprites, their interactions, and the positions
they have in different games and levels. The contribution of this work is stimulating the generation of new ideas by automatically exploring the design
spaces of games, and provide solutions that the designer does
not need to generate from scratch.
Perhaps the main limitation of the  recommender approach
is that it would seem to encourage  sameness and quasi-plagiarism. But, as discussed above,  users  can tune their “originality” by balancing their choices between high and low confidence recommendations. In the future,  it would be interesting
to complement the current data-driven recommendations with
recommendations based e.g. on evolutionary search. 
A
preliminary and informal study showed that the mixed design
initiative was well employed by the  users, with them having
half of their game design algorithmically suggested.
We hope this work will help push the use of recommender systems beyond the e-commerce  field, and that it can be
used more as a method of (game) design assistance.

\section*{Acknowledgements}
Tiago Machado is supported by the Conselho Nacional de Desenvolvimento Cient\'ifico e Tecnol\'ogico (CNPQ), under the Science without Borders scholarship 202859/2015-0. We also want to thank to our intern students and collaborators, Angela Wang, Katherine LosCalzo, Katalina Park, and ZhongHeng Li.

\bibliographystyle{IEEEtran}
\bibliography{conference_041818}

\end{document}